\documentclass[runningheads]{llncs}
\usepackage{graphicx}
\usepackage{subfigure}
\usepackage[misc]{ifsym}
\begin{document}
\title{A Hierarchical Optimizer for Recommendation System Based on Shortest Path Algorithm}
%
%

\author{Jiacheng Dai \and Zhifeng Jia \and Xiaofeng Gao\thanks{Xiaofeng Gao is the corresponding author.} \and Guihai Chen}
\authorrunning{J. Dai et al.}
%
\institute{Shanghai Key Laboratory of Scalable Computing and Systems,\\
	 Department of Computer Science and Engineering, Shanghai Jiao Tong University,\\
	 Shanghai 200240, China\\ 
\email{\{daijiacheng,fergusjia\}@sjtu.edu.cn}, \email{\{gao-xf,gchen\}@cs.sjtu.edu.cn}}
\maketitle              

\vspace{-20pt}
\begin{abstract}
Top-k Nearest Geosocial Keyword (T-kNGK) query on geosocial network is defined to give users $k$ recommendations based on some keywords and designated spatial range, and can be realized by shortest path algorithms. However, shortest path algorithm cannot provide convincing recommendations, so we design a hierarchical optimizer consisting of classifiers and a constant optimizer to optimize the result by some features of the service providers.

\vspace{-8pt}
\end{abstract}

\vspace{-30pt}
\section{Problem Statement}

\vspace{-5pt}

\begin{figure}[t]
	\begin{minipage}[t]{0.5\linewidth}
		\centering   
		\includegraphics[width=2.3in]{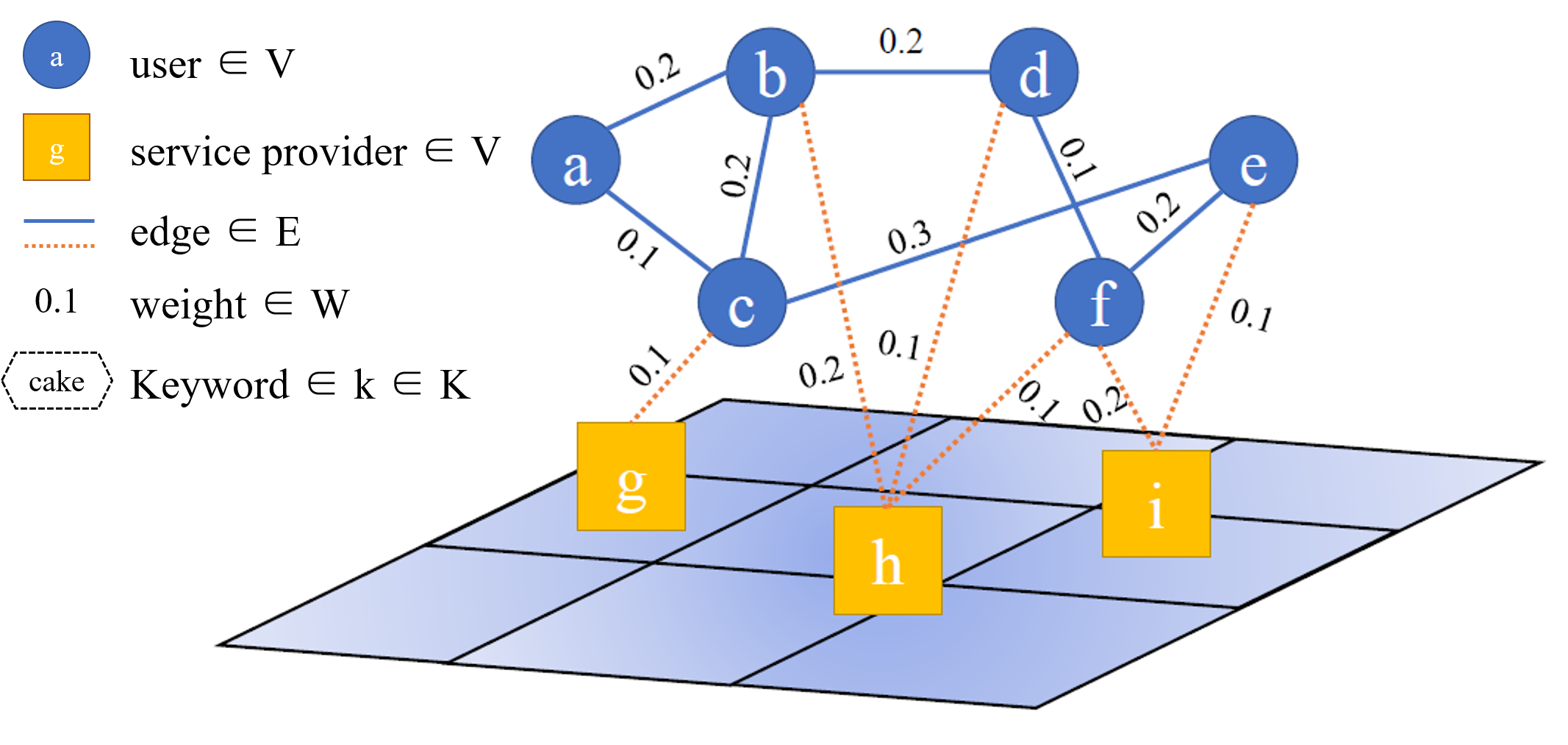}   
		\vspace{-10pt}
		\caption{An instance of geosocial network}   
		\label{geo}   
	\end{minipage}%
	\begin{minipage}[t]{0.5\linewidth}   
		\centering   
		\includegraphics[width=1.95in]{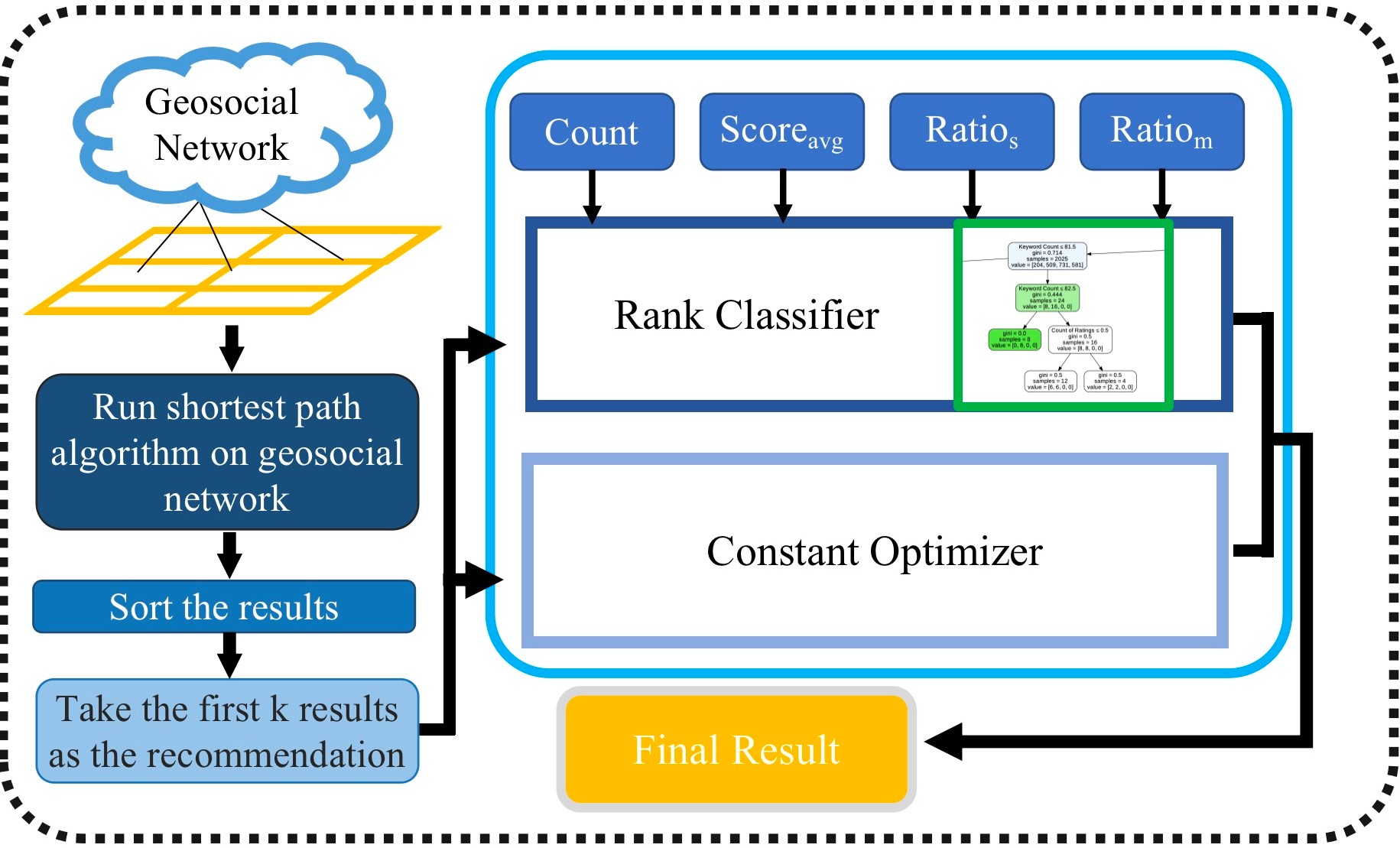}   
		\vspace{-10pt}
		\caption{The optimized T-kNGK query}   
		\label{tkngk}   
	\end{minipage}   
\end{figure}

Top-k Nearest Geosocial Keyword Search Query (T-kNGK) works on a geosocial network. A T-kNGK query's task is to recommend $k$ service providers (SP's) that best meet the user's requirement (keywords and location). Shortest path algorithms are used to solve this problem. We simply take the length of the shortest path between user $u$ and SP $v$ as the basis for recommendation. A geosocial network \cite{sun} is a weighted undirected graph $G = (V,E,W,K,L)$. We give a simple instance Figure \ref{geo} to illustrate its structure and contents. The weight of an edge shows the intimacy between users or rating for SPs (mapped to $[0,1]$). Each SP $v$ has a keyword set $k_v \in K$ and location $l_v \in L$. The process of a T-kNGK query is shown in Figure \ref{tkngk} (together with the Hierarchical Optimizer). The defect of T-kNGK query is that it seems to be unconvincing since it only references one comment (one path) due to shortest path algorithm. Therefore, we should optimize the results of T-kNGK query to enhance reliability and avoid extreme bad cases.

\vspace{-10pt}

\section{Hierarchical Optimizer}

\vspace{-5pt}

\subsubsection{Constant Optimizer}
We try to let the SP's with more comments get higher score and reduce the score of SP's who have few comments. We define multiplier $\alpha$ to calculate the new score $Score_c$. $\alpha_{s_i}=1+\frac{1}{\beta} (\frac{count_{s_i}-average}{average-min_{1\leq j \leq n}\{count_{s_j}\}})^{\gamma}$ when $count_{s_i}<average$ and \label{alpha2}
$\alpha_{s_i}=1+\frac{1}{\beta} (\frac{count_{s_i}-average}{max_{1\leq j \leq n}\{count_{s_j}\}-average})^{\gamma}$ when $count_{s_i}\geq average$, where $average=\frac{\Sigma_{j=1}^n count_{s_j}}{n}$, $\beta$ and $\gamma$ are adjustable parameters and $count_{s_i}$ is the number of comments of SP $s_i$. Finally, $Score_c = \alpha_{s_i} \times rating_{s_i}$.

\subsubsection{Rank Classifier}
We defined four features to train a classifier which ranks the SP's from 1 to 5: ``Matched Keyword Ratio'' $Ratio_m = \frac{|k_q \cap k_{s_i}|}{|k_q|}$, ``Specific Keyword Ratio'' $Ratio_s = \frac{|k_q \cap k_{s_i}|}{|k_{s_i}|}$, ``Count of Ratings'' $Count_{s_i}$, ``Average Score'' $Score_{avg_{s_i}}$.

In the Hierarchical Optimzer, we first use Rank Classifier to rank the SP's we get from the shortest path algorithm and sort them by their rank. Then we use Constant Optimizer to calculate $Score_c$ and sort the SP's of the same rank by $Score_c$. Thus, we got the optimized results.

\vspace{-10pt}

\section{Experiments and Results}

\vspace{-8pt}

\begin{figure}[h]
	\centering
	\includegraphics[width=4.6in]{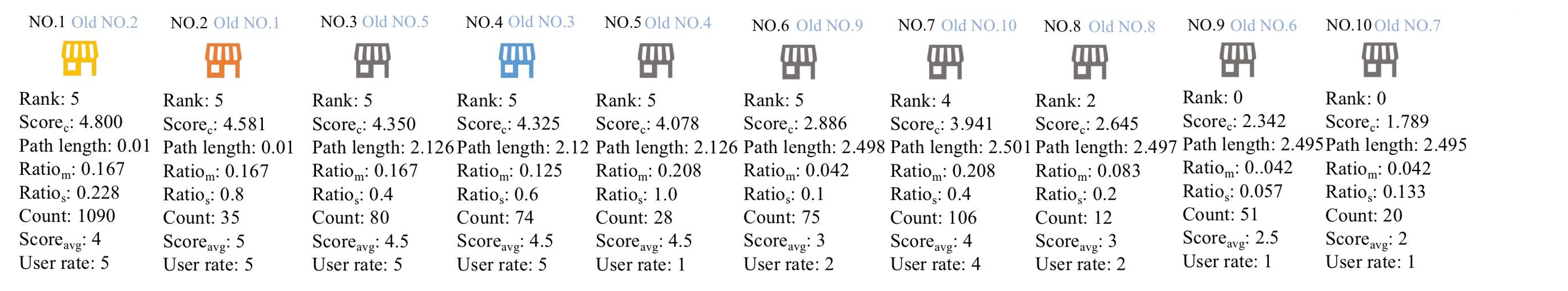}
	\vspace{-10pt}
	\caption{The result of optimization}
	\label{res}
\end{figure}

We use Yelp dataset (over 3 GB and has over 6.5 million reviews) which contains all the data we need in the T-kNGK query and the hierarchical optimizer. For Constant Optimizer, we set $\beta = 5$ and $\gamma = 2$. For Rank Classifier, we choose random forest with accuracy of 82\%. We use 80\% of the data to train the model and 20\% to test the model. First, we run shortest path algorithm on the geosocial network built on Yelp dataset and get some raw result. Then we use our Hierarchical Optimizer to re-order the raw results and get ideal result. One of our optimized result is shown in Figure \ref{res}.

\vspace{-10pt}

\vspace{-30pt}

\subsubsection{Acknowledgments.}This work was supported by the National Key R\&D Program of China [2018YFB1004703]; the National Natural Science Foundation of China [61872238, 61672353]; the Shanghai Science and Technology Fund [17510740200]; the Huawei Innovation Research Program [HO2018085286]; the State Key Laboratory of Air Traffic Management System and Technology [SKLAT
\\
M20180X]; and the Tencent Social Ads Rhino-Bird Focused Research Program.

\vspace{-12pt}
%
%
%

\begin{thebibliography}{8}

\vspace{-9pt}
\bibitem{sun}
Y. Sun, N. Pasumarthy, and M. Sarwat, “On Evaluating Social Proximity-Aware
Spatial Range Queries,” in IEEE International Conference on Mobile Data Man-
agement (MDM), KAIST, Taejeon, South Korea, 2017, pp. 72–81.

\end{thebibliography}
%

\end{document}